\documentstyle[amsmath,amssymb,graphicx]{article}

\def\be{\begin{eqnarray}}
\def\ee{\end{eqnarray}}
\def\nn{\nonumber}

\def\l[{\phantom.[}

\hoffset= - 1.1in         % switch off for draft style
\begin{document}

\hfill ITEP/TH-11/15

\hfill IITP/TH-4/15

\bigskip

\centerline{\Large{
Factorization of colored knot polynomials at roots of unity
}}

\bigskip

\centerline{{\bf Ya.Kononov$^d$, A.Morozov$^{a,b,c}$}}

\bigskip

{\footnotesize
\centerline{{\small
$^a$ ITEP, Moscow 117218, Russia}}

\centerline{{\small
$^b$ National Research Nuclear University MEPhI, Moscow 115409, Russia
}}

\centerline{{\small
$^c$ Institute for Information Transmission Problems, Moscow 127994, Russia
}}

\centerline{{\small
$^d$ Higher School of Economics, Math Department, Moscow, 117312, Russia
}}
}

\bigskip

\bigskip

\centerline{ABSTRACT}

\bigskip

{\footnotesize
From analysis of a big variety of different knots we conclude that
at $q$, which is an $2m$-th root of unity, $q^{2m}=1$,
HOMFLY polynomials in symmetric representations $[r]$ satisfy recursion
identity:  $H_{r+m} = H_r\cdot H_m$ for any $A$,
which is a generalization of the property $H_r = H_1^r$ for special polynomials
at $m=1$.
We conjecture a further generalization to arbitrary representation $R$,
which, however, is checked only for torus knots.
Next, Kashaev polynomial, which arises from $H_R$ at $q^2=e^{2\pi i/|R|}$,
turns equal to the special polynomial with $A$ substituted by $A^{|R|}$,
provided $R$ is a single-hook representations (including arbitrary symmetric) --
what provides a $q-A$ dual to the similar property of Alexander polynomial.
All this implies non-trivial relations for the coefficients of the differential
expansions, which are believed to provide reasonable coordinates in
the space of knots -- existence of such universal relations means that these variables
are still not unconstrained.
}

\bigskip

\bigskip

Knot polynomials \cite{knotpols,katlas} are Wilson loop averages in Chern-Simons theory \cite{CS}
and their study provides important knowledge and intuition for understanding
the properties of gauge-invariant observables in generic Yang-Mills theory.
Since Chern-Simons theory is topological, the space-time dependence is completely
decoupled and one can extract pure information about the representation (color) dependence.
However, the problem of calculating colored HOMFLY polynomials
\be
H_R^{\cal K}(A,q^2) = \left< {\rm Tr}_R \ \text{P}\exp\left(\oint_{\cal K} {\cal A}\right)\right>^{CS}
\ee
with the gauge group $Sl(N)$ and coupling constant $g$ converted into
$q^2=\exp\left(\frac{2\pi i}{g+N}\right)$
and $A=q^N$,
turned to be highly non-trivial.
Only recently considerable advances were achieved in \cite{gmmms,mmmrs},
based on decades of the previous work \cite{ASch}-\cite{knotslast},
opening a possibility to look for properties,
that are valid universally, i.e. for arbitrary knots.
In \cite{KoModef} we showed, how this new information leads to immediate breakthrough in the theory
of differential expansions \cite{IMMMfe,evo,arthdiff}.
These expansions provide a non-trivial knot-dependent "quantization" of the
archetypical factorization property \cite{LP}-\cite{Anton}
\be
\sigma_R^{\cal K}(A)= \Big(\sigma_{[1]}^{\cal K}(A)\Big)^{|R|}
\label{fasigma}
\ee
of the {\it special} polynomials $\sigma_R(A)= H_R(q^2=1,A)$, which are restriction of HOMFLY
to $q^2=1$, i.e. a kind of their large-$N$ limit.
Differential expansion substitutes {\it factorization} at $q=1$ by {\it expansion} at $q\neq 1$,
which, however, contains finitely many terms with their own pronounced factorization properties.
They are best studied for symmetric representation $R=[r]$:
\be
{\cal H}_r^{\cal K}(A,q^2|h) \ =\ 1\ +\  \ \sum_{s=1}^r \
\frac{[r]!}{[s]![r-s]!}\cdot h^s \, G_s^{\cal K}(A,q)
\cdot\{A/q\}\cdot \prod_{j=0}^{s-1} \{Aq^{r+j}\}
\label{diffe}
\ee
We introduced here and additional parameter $h$, distinguishing the "level" of differential expansion.
HOMFLY polynomial itself arises at $h=1$:
\be
H_r^{\cal K}(A,q^2) ={\cal H}_r^{\cal K}(A,q^2|h=1)
\ee

\newpage

\section{Relations at the roots of unity. Symmetric representations}

In this paper we study {\it another} generalization of (\ref{fasigma}),
which preserves its {\it factorized} form, but is instead true only at particular
values of $q$ -- namely, at roots of unity.
It turns out that for $q^{2m}=1$
\be
\boxed{
\left. H_{r+m}^{\cal K} = H_{r}^{\cal K}\cdot H_m^{\cal K} \ \right|_{q^{2m}=1}}
 \label{facrpm}
\ee
where $H_r^{\cal K} = H_{[r]}^{\cal K}$ is HOMFLY polynomial in the totally symmetric representation $[r]$
(i.e. Young diagram is a single line of length $r$).

\bigskip

{\footnotesize
As an example take the knot ${\cal K} =  6_2$ from the Rolfsen table at $m=2$.
For $q=\pm 1$ we have (\ref{fasigma}) with $\sigma_{[1]}^{6_2} = \left.H_1^{6_2}\right|_{q=\pm 1}
= \frac{1-2A^2+2A^4}{A^4} $.
At the other two roots $q=\pm i$ we get:
\be
\left.H_1^{6_2}\right|_{q=\pm i} = -\frac{2 A^4+6 A^2+3}{A^4} \nn\\
\left.H_2^{6_2}\right|_{q=\pm i} = \frac{2 A^8-2 A^4+1}{A^8} \nn\\
\left.H_3^{6_2}\right|_{q=\pm i} = -\frac{\left(2 A^4+6 A^2+3\right) \left(2 A^8-2 A^4+1\right)}{A^{12}}\nn\\
\left.H_4^{6_2}\right|_{q=\pm i} = \frac{\left(2 A^8-2 A^4+1\right)^2}{A^{16}}\nn\\
\left.H_5^{6_2}\right|_{q=\pm i} = -\frac{\left(2 A^4+6 A^2+3\right) \left(2 A^8-2 A^4+1\right)^2}{A^{20}}\nn\\
\left.H_6^{6_2}\right|_{q=\pm i} = \frac{\left(2 A^8-2 A^4+1\right)^3}{A^{24}}\nn\\ \nn \\
\ldots \nn
\ee
also in full accordance with (\ref{facrpm}).
}

\bigskip

Original (\ref{fasigma}) is now a particular case of (\ref{facrpm}) with $m=1$.
Since transposition of Young diagram $R\longrightarrow \tilde R$
is equivalent to the substitution $q\longrightarrow q^{-1}$
\cite{DMMSS},
\be
H_{\tilde R}^{\cal K}(q^2,A) = H_R^{\cal K}\left(\frac{1}{q^2},\,A\right)
\label{transH}
\ee
the same recursion holds for totally antisymmetric representations $[1^r]$
(Young diagram is a column of length $r$):
\be
\left. H_{[1^{r+m}]}^{\cal K} = H_{[1^r]}^{\cal K}\cdot H_{[1^m]}^{\cal K}\ \right|_{q^{2m}=1}
\ee
In fact, (\ref{facrpm}) is equivalent to a more symmetric statement:
\be
\boxed{
\Big(H_{r+m}^{\cal K} - H_{r}^{\cal K}\cdot H_m^{\cal K}\Big) \ \ \vdots\ \  \
\frac{\{q^r\}\{q^m\}}{\{q^{{\rm gcd}(r,m)}\}}
\cdot\{A/q\}
}
%= \{q\} \{A/q\} \prod_{i}[k_i]
\ \ \ \ \ \ \ \Longleftrightarrow \ \ \ \ \ \ \
\boxed{
\Big( H_{[1^{r+m}]}^{\cal K} - H_{[1^r]}^{\cal K}\cdot H_{[1^m]}^{\cal K}\Big) \ \ \vdots\ \  \
\frac{\{q^r\}\{q^m\}}{\{q^{{\rm gcd}(r,m)}\}}
\cdot\{Aq\}
}
\label{facrmm}
\ee
where ${\rm gcd}(r,m)$ is the greatest common divisor of $r$ and $m$,
%$k_i$ are the simple factors in the product $mn=\prod_i k_i^{s_i}$
and $\{x\}=x-x^{-1}$,
so that the quantum number $[p] = \{q^p\}/\{q\}$ and
for coprime $r$ and $m$ the r.h.s. is just $[r][m]\cdot \{q\}\{A/q\}$.
The statement is that the r.h.s. factors out from the difference at the l.h.s. at any $A$ and $q$.
%The point is that multiplicities $s_i$ drop out of (\ref{facrmm}), only if $r$ and $m$
%are coprime, we can write just $[r][m]$ at the r.h.s.

It can be interpreted as one more property of the differential expansion (\ref{diffe}):
\begin{multline}
{\cal H}_{r+m}^{\cal K} - {\cal H}_{r}^{\cal K}\cdot {\cal H}_m^{\cal K} =  \\
= \sum_{s=1}^{r+m} \ h^s\cdot \left(\frac{[r+m]!}{[s]![r+m-s]!}\prod_{j=0}^{s-1}\{Aq^{r+m+j}\}
-\frac{[r]!}{[s]![r-s]!}\prod_{j=0}^{s-1}\{Aq^{r+j}\}
- \frac{[m]!}{[s]![m-s]!}\prod_{j=0}^{s-1}\{Aq^{m+j}\}\right)\cdot\{A/q\}\cdot G_{s}^{\cal K}
+  \\
+ \sum_{s',s''=1} h^{s'+s''}\cdot\frac{[r]![m]!}{[s']![s'']![r-s']![m-s'']!}
\prod_{j=0}^{s'-1}\{Aq^{r+j}\} \prod_{j=0}^{s''-1}\{Aq^{m+j}\}
\cdot \{A/q\}^2 \cdot G_{s'}^{\cal K}G_{s''}^{\cal K}
\label{prooffromdiffe}
\end{multline}
Many  terms in these sums are immediately proportional to the r.h.s. of (\ref{facrmm}), but not all.
Even the factor $\{q\}$ at $m=1$ is not immediately obvious from (\ref{prooffromdiffe}), but
at $q=\pm 1$ identity (\ref{fasigma}) can be additionally used:
\be
(\ref{fasigma})  \ \ \ \ \Longrightarrow \ \ \ \ \left. G_s = \{A\}^{s-1}\cdot G_1^s\ \right|_{q^2=1}
\ee
Still it turns out -- and this is a highly non-trivial additional fact --
 that {\bf proportionality to the r.h.s. of (\ref{facrmm})
holds independently} at each level $s=s'+s''$, i.e. {\bf in each order of the $h$-expansion},
thus {\it enhancing}
(\ref{facrmm}) to a whole set of quadratic restrictions on the values of $G_s^{\cal K}$
at roots of unity:
\be
%\left. \{A\}\cdot G_3^{\cal K} = \{Aq^2\}\{A/q\}\cdot G_2^{\cal K}G_1^{\cal K} \ \right|_{q^4=1}  \nn \\
\left.  G_3^{\cal K} = q^2\cdot \{A/q\}\cdot G_2^{\cal K}G_1^{\cal K} \ \right|_{q^4=1}  \nn \\
\ldots \ \ \ \ \ \ \ \ \ \ \ \ \ \ \ \ \ \ \ \ \ \ \ \ \ \ \ \ \ \ \ \ \nn \\
\boxed{
\left. \prod_{j=0}^{m-1}\{Aq^{2r+j}\}\cdot G_{r+m}^{\cal K} = q^{rm}\cdot \{A/q\}
\prod_{j=0}^{m-1}\{Aq^j\}\cdot G_r^{\cal K}G_m^{\cal K} \ \right|_{q^{2m}=1}
}
\nn \\
\ldots  \ \ \ \ \ \ \ \ \ \ \ \ \ \ \ \ \ \ \ \ \ \ \ \ \ \ \ \ \ \ \ \
\label{relsforG}
\ee
%Especially nice
The nicely-looking
relation in the box arises at the order $h^{m+r}$, but it does not exhaust the set of relations:
there are many more, arising at smaller powers of $h$ in between $\text{max}(r,m)$ and $r+m$,
but they look less elegant.

A useful corollary of (\ref{relsforG}) is
\be
G_{ms} \ + \ q^{m^2 s (s-1)/2}  \cdot \{A/q\}^{s-1}\cdot  (G_s)^m
\ \ \ \vdots \ \ \ q^{2m}-1
\ee

\section{Beyond symmetric representations}

To really be a {\it generalization} of (\ref{fasigma}), relations like (\ref{facrpm})
should hold for arbitrary representations $R$, not only symmetric.
Indeed it looks like there are plenty of them, and they continue to respect
% non-symmetric case there are a lot of similar relations respecting
the grading by the level (number of boxes in Young diagram) --
all such relations at special values of $q$  are homogeneous in this grading.
However it is difficult to find the reliable general rule.
In this section we describe the relations at low levels $|R|$ and formulate a plausible
general conjecture.

\subsection{Extension to $[21]$}

Beyond symmetric representations the story is more complicated, because the analogue
of differential expansion (\ref{diffe}) is still unknown.
Moreover, not much is known about the non-symmetrically HOMFLY at all,
even examples are restricted mostly to torus knots.
The latest breakthrough in \cite{mmmrs} provides  answers for rather general knots,
but only for $R=[21]$.
Still, this very restricted result allows us to move further.

From the data, obtained on the lines of \cite{mmmrs} we conclude empirically that
the relevant generalization of (\ref{fasigma}) for $R=[21]$ is to arbitrary roots of order $6$:
\be
\boxed{\left. H_{[21]}^{\cal K} = H_{[3]}^{\cal K} = H_{[111]}^{\cal K}\ \right|_{q^6=1}} \ \ \
\Longleftarrow \ \ \
\Big(H_{[21]}^{\cal K}-H_{[3]}^{\cal K}\Big) \ \ \vdots \ \ \{q^3\}\{A\}
\label{fac21r6}
\ee
Moreover, the second equality in (\ref{fac21r6}) has its own generalization:
\be
\boxed{
\Big(H_{[r]}^{\cal K} - H_{[1^r]}^{\cal K}\Big) \ \ \vdots \ \ [r][r-1]\{q\}\{A\}
} \ \ \ \Longrightarrow \ \ \ \left.H_{[r]}=H_{[1^r]}\right|_{q^{2r}=1}, \ \ \ \
\left.H_{[r]}=H_{[1^r]}\right|_{q^{2r-2}=1}
\label{dir1r}
\ee

Unfortunately (\ref{fac21r6}) is all what we can check at this moment for rather general knots.
In order to move further in non-symmetric case, we need to take a more risky road.

\subsection{Implications from torus knots
\label{torus}}

After the very phenomenon is revealed from analysis of a rather general data,
it can be further investigated on a far more restricted data field.
Namely, if we believe/assume that there are {\it universal} relations
between colored HOMFLY at roots of unity, i.e. valid for {\it all} knots,
their concrete shape can be found by looking at particular knot families.
Reliability of such statements is, of course, restricted, and what we get in this
way are just {\it conjectures}.
Still they can be brought to a relatively nice form and it is plausible
that they are {\it universally} true.

Torus knots provide a natural family to look at,
because this is the only one, where colored HOMFLY are available
in arbitrary representation.
This is because torus knots are more representation-theory than topological objects,
and one should be very careful when extending observations made for this family
to generic case -- still we believe that conjectures below have good chances to
be universally reliable.

HOMFLY for torus knot $[m,n]$ is given by the Rosso-Jones formula \cite{RJ,LP,DMMSS}
\be
H_R^{[m,n]} = \sum_{|Q| = m |R|} c_R^Q q^{2n/m \kappa_Q} \chi_Q/\chi_R,
\ee
where $\kappa_Q$ is the content of the diagram, and $c_R^Q$ are matrix elements
of the so-called Adams endomorphism in the basis of Schur polynomials.

\subsection{Conjecture}

From the study of torus knots we make
a {\bf conjecture}, which is presumably valid for all knots:
\be
\boxed{\
\left.H_{R+M}^{\cal K}\ \stackrel{?}{=}\ H_m^{\cal K}\cdot H_R^{\cal K}\right|_{q^{2m}=1}
\ \ \ \ \ \forall \ \text{ connected skew diagram}\ M \ \text{of width one with}\ |M|=m
\ }
\label{Rpm}
\ee
provided both $R$ and $R+M$ are Young diagrams.
%, i.e. inequalities are true in
%$R=\{r_1\geq r_2 \geq \ldots \geq r_{l_R}> 0\}$ and $R+m_j=\{r_1\geq r_2\geq\ldots
%\geq r_{j-1}\geq r_j+m\geq r_{j+1}\geq \ldots\geq r_{l_R}> 0\}$.
The following picture is an explanation of what we mean by  $R+M$:

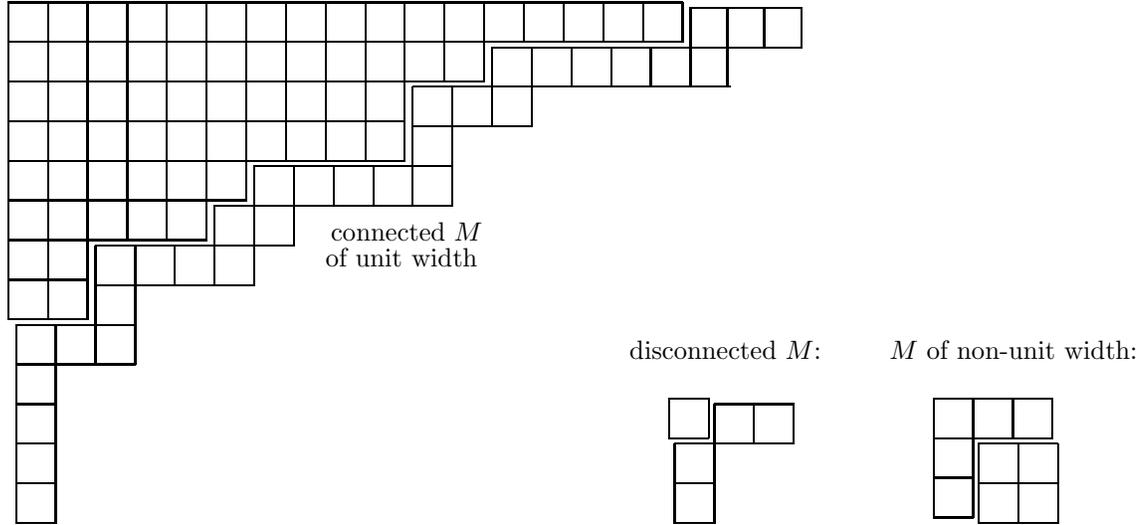
\begin{picture}(300,225)(-50,-205)
\put(0,0){\line(1,0){255}}
\put(0,-15){\line(1,0){255}}
\put(0,-30){\line(1,0){180}}
\put(0,-45){\line(1,0){150}}
\put(0,-60){\line(1,0){150}}
\put(0,-75){\line(1,0){90}}
\put(0,-90){\line(1,0){75}}
\put(0,-105){\line(1,0){30}}
\put(0,-120){\line(1,0){30}}
\put(0,0){\line(0,-1){120}}
\put(15,0){\line(0,-1){120}}
\put(30,0){\line(0,-1){120}}
\put(45,0){\line(0,-1){90}}
\put(60,0){\line(0,-1){90}}
\put(75,0){\line(0,-1){90}}
\put(90,0){\line(0,-1){75}}
\put(105,0){\line(0,-1){60}}
\put(120,0){\line(0,-1){60}}
\put(135,0){\line(0,-1){60}}
\put(150,0){\line(0,-1){60}}
\put(165,0){\line(0,-1){30}}
\put(180,0){\line(0,-1){30}}
\put(195,0){\line(0,-1){15}}
\put(210,0){\line(0,-1){15}}
\put(225,0){\line(0,-1){15}}
\put(240,0){\line(0,-1){15}}
\put(255,0){\line(0,-1){15}}
\put(258,-2){\line(1,0){42}}
%\put(258,-15){\line(1,0){42}}
\put(183,-17){\line(1,0){117}}
\put(153,-32){\line(1,0){120}}
\put(153,-47){\line(1,0){45}}
\put(93,-62){\line(1,0){75}}
\put(78,-77){\line(1,0){90}}
\put(33,-92){\line(1,0){75}}
\put(33,-107){\line(1,0){60}}
\put(3,-122){\line(1,0){45}}
\put(3,-137){\line(1,0){45}}
\put(3,-152){\line(1,0){15}}
\put(3,-167){\line(1,0){15}}
\put(3,-182){\line(1,0){15}}
\put(3,-197){\line(1,0){15}}
\put(3,-122){\line(0,-1){75}}
\put(18,-122){\line(0,-1){75}}
\put(33,-92){\line(0,-1){45}}
\put(48,-92){\line(0,-1){45}}
\put(63,-92){\line(0,-1){15}}
\put(78,-77){\line(0,-1){30}}
\put(93,-62){\line(0,-1){45}}
\put(108,-62){\line(0,-1){30}}
\put(123,-62){\line(0,-1){15}}
\put(138,-62){\line(0,-1){15}}
\put(153,-32){\line(0,-1){45}}
\put(168,-32){\line(0,-1){45}}
\put(183,-17){\line(0,-1){30}}
\put(198,-17){\line(0,-1){30}}
\put(213,-17){\line(0,-1){15}}
\put(228,-17){\line(0,-1){15}}
\put(243,-17){\line(0,-1){15}}
\put(258,-2){\line(0,-1){30}}
\put(272,-2){\line(0,-1){30}}
\put(286,-2){\line(0,-1){15}}
\put(300,-2){\line(0,-1){15}}
\put(250,-150){\line(1,0){15}}
\put(250,-165){\line(1,0){15}}
\put(250,-150){\line(0,-1){15}}
\put(265,-150){\line(0,-1){15}}
\put(267,-152){\line(1,0){30}}
\put(252,-167){\line(1,0){45}}
\put(252,-182){\line(1,0){15}}
\put(252,-197){\line(1,0){15}}
\put(252,-167){\line(0,-1){30}}
\put(267,-152){\line(0,-1){45}}
\put(282,-152){\line(0,-1){15}}
\put(297,-152){\line(0,-1){15}}
\put(122,-90){\mbox{connected $M$}}
\put(120,-100){\mbox{of unit width}}
\put(235,-135){\mbox{disconnected $M$:}}
\put(330,-135){\mbox{ $M$ of non-unit width:}}
\put(350,-150){\line(1,0){45}}
\put(350,-165){\line(1,0){45}}
\put(350,-180){\line(1,0){15}}
\put(350,-195){\line(1,0){15}}
\put(350,-150){\line(0,-1){45}}
\put(365,-150){\line(0,-1){45}}
\put(380,-150){\line(0,-1){15}}
\put(395,-150){\line(0,-1){15}}
\put(367,-167){\line(1,0){30}}
\put(367,-182){\line(1,0){30}}
\put(367,-197){\line(1,0){30}}
\put(367,-167){\line(0,-1){30}}
\put(382,-167){\line(0,-1){30}}
\put(397,-167){\line(0,-1){30}}
\end{picture}

Both conditions in (\ref{Rpm}) are necessary.
The simple counterexamples are:

\bigskip

$\bullet$ $\left.H_{[311]}\neq H_4 H_{[1]}\neq H_{[2]}^2H_{[1]}\right|_{q^8=1}$ --
connectedness is indeed needed
and

$\bullet$ $\left.H_{[333]}\neq H_{4}H_{[311]}\neq H_{[22]}H_{[311]}\right|_{q^8=1}$ --
width $2$ is too much

\bigskip

%Note that also $\left.H_{[311]}\neq H_{[2]}^2H_{[1]}\right|_{q^8=1}$
%and $\left.H_{[333]}\neq H_{[22]}H_{[311]}\right|_{q^8=1}$.

Note that (\ref{dir1r}) implies that  $H_{[m]}=H_{[1^m]}$ at $q^{2m}=1$ -- thus both can
play the role of $H_m$ in (\ref{Rpm}).
Alternatively one can say that (\ref{dir1r}) is a particular case of the more general conjecture
(\ref{Rpm}) -- at the moment the difference is that the former is checked for all kinds
of knots, while the latter -- only for torus ones.

\subsection{Examples of (\ref{dir1r})+(\ref{Rpm})
\label{Exa}}

To illustrate the implications of (\ref{Rpm}) -- or, if one prefers, the evidence in support of it --
we now provide a few simple examples.
We checked that all these relations are indeed true for a big variety of torus knots --
and, as explained in above sec.\ref{torus} we believe that they hold for all other knots,
though this remains to be checked when the corresponding colored polynomials become available.

\subsubsection{ Factorization at $q^4=1$}

In this case there are two options for $M$: $[2]$ and $[11]$ --
a line and a column on length $m=2$.
Adding these two elements to various $R$ we obtain from (\ref{Rpm}):

\bigskip

$\bullet$ Level $|R|+|M|=3$:  \ \ \ $\Box \sim \Box \Box\Box \sim {\Box\atop{\Box\atop\Box}}$
\be
H_3 = H_{[111]} = H_1 H_2 \neq H_{[21]}
\ee
Coincidence between $H_3$ and $H_{[111]}$ is also a corollary of the factor $[r-1]$ in (\ref{dir1r}).

\bigskip

$\bullet$ Level 4:

Here we have two parent diagrams $R=[2]$ and $R=[11]$, but their HOMFLY
are identical at $q^4=1$, $H_{[2]}=H_{[11]}$ due to (\ref{dir1r}). Thus from (\ref{Rpm})
\be
H_Q = H_2^2 \ \ \ \ \ \forall Q ,\ \ \ |Q| = 4
\ee

$\bullet$ Level 5:
\be
H_5 = H_{[32]} = H_{[311]} = H_{[221]} = H_{[1^5]} = H_1 H_2^2 \nn\\
H_{[41]} = H_{[2111]} = H_{[21]} H_2
\ee

$\bullet$ Level 6:
\be
H_Q = H_2^3 \ \ \ \ \ \forall Q \neq {[321]} \ \ \ |Q| = 6
\ee

$\bullet$ Level 7:
\be
H_7 = H_{[52]} = H_{[511]} = H_{[421]} = H_{[331]} = H_{[322]}
= H_{[3211]} = H_{[311111]} = H_{[22111]} = H_{[1^7]} = H_1 H_2^2; \nn\\
H_{[61]} = H_{[43]} = H_{[4111]} = H_{[2221]}=H_{[211111]} = H_{[21]} H_2^2
\ee

$\bullet$ Level 8:
\begin{multline}
H_8 = H_{[71]} = H_{[62]} = H_{[611]} = H_{[53]} = H_{[5111]} = H_{[44]}
= H_{[431]} = H_{[422]} = H_{[4211]} = H_{[41111]} = \nn \\
=H_{[332]} = H_{[3311]} = H_{[3221]} = H_{[311111]} = H_{[2222]} = H_{[22211]}
= H_{[221111]} = H_{[2111111]} = H_{[1^8]} = H_2^4;
\end{multline}
\be
H_{[521]} = H_{[32111]} = H_{[321]} H_2
\ee

\subsubsection{ Factorization at $q^6=1$}

First of all, we have universally valid relation (\ref{transH}),
which is also applicable at roots of unity, where inversion
gets equivalent to complex conjugation: $q^{-1}=q^*$. Thus,

\bigskip

$\bullet$ Level 2:
\be
H_2 = H_{[11]}^*
\ee

Note, however, that $*$ is not supposed to act on $A$: $A^*=A$

\bigskip

Next, from (\ref{fac21r6}):

\bigskip

$\bullet$ Level 3:
\be
H_3 = H_{[21]} = H_{[111]}
\ee

Now we have for options four skew diagrams $M$ of length $|M|=3$:  \
$[3]$, $[111]$, $[21]$ and $\overline{[21]}$
where the last one is an upside-down version
$\overline{[21]}={\ \ \Box\atop{\Box\Box}}$ of $[21]={{\Box\Box}\atop\Box\ \ }$\
Adding them to various $R$ we obtain the following ''orbits'':

\bigskip

$\bullet$ Level 4:
\be
H_4 = {H_{[22]}} = H_{[1111]} = H_1 H_3 \nn \\
H_{[31]} = H_{[211]}^*
\ee

Coincidence between $H_4$ and $H_{[1111]}$ follows directly from (\ref{dir1r}),
but it is also an implication of (\ref{Rpm}) -- when we add either $M=[3]$
or $M=[111]$ to $R=[1]$. If $M=\overline{[21]}$ is added instead, we get $H_{[22]}$.
However, there is no way to get $H_{[31]}\stackrel{(\ref{transH})}{=}H_{[211]}^*$
and it is indeed independent.

\bigskip

$\bullet$ Level 5:
\be
H_5 = H_{[221]} = H_{[2111]} = H_2 H_3 \nn\\
H_{[41]} = H_{[3,2]} = H_{[11111]} = H_{[11]} H_3
\ee

$\bullet$ Level 6:
\be
H_6 = H_{[51]} = H_{[411]} = H_{[33]} = H_{[321]} = H_{[3111]} = H_{[222]}
= H_{[21111]} = H_{[111111]} = H_3^2\nn\\
H_{[42]} = H_{[2211]}^*
\ee

$\bullet$ Level 7:

\be
H_7 = H_{[52]} = H_{[43]} = H_{[421]} = H_{[4111]} = H_{[3211]}
= H_{[2221]} = H_{[22111]} = H_{[11111111]} = H_3^2 H_1 \nn\\
H_{[61]} = H_{[322]} = H_{[31111]} = H_3 H_{[31]}\nn \\
H_{[511]} = H_{[331]} = H_{[211111]} = H_{[211]} H_3
\ee

$\bullet$ Level 8:
\be
H_8 = H_{[53]} = H_{[521]} = H_{[5111]} = H_{[431]} = H_{[3311]}
= H_{[2222]} = H_{[221111]} = H_{[2111111]} = H_3^2 H_2 \nn\\
H_{[71]} = H_{[62]} = H_{[44]} = H_{[422]} = H_{[41111]} = H_{[3221]} = H_{[32111]}
= H_{[22211]} = H_{[11111111]} = H_3^2 H_{[11]} \nn \\
H_{[611]} = H_{[332]} = H_{[311111]} = H_3 H_{[311]} \nn\\
H_{[4211]}
\ee

\subsubsection{Factorizations at $q^8=1$}

\noindent

$\bullet$ Level 4:
\be
H_4 = H_{[31]} = H_{[211]} = H_{[1111]} \nn \\
H_{[22]}
\ee

$\bullet$ Level 5:
\be
H_5 = H_{[32]} = H_{[221]} = H_{[11111]} = H_4 H_1 \nn \\
H_{[41]} = H_{[2111]}^* \nn\\
H_{[311]}
\ee

$\bullet$ Level 6:
\be
H_6 = H_{[33]} = H_{[2211]} = H_{[21111]} = H_4 H_2 \nn \\
H_{[51]} = H_{[42]} = H_{[222]} = H_{[111111]} = H_4 H_{[11]} \nn\\
H_{[411]} = H_{[3111]}^* \nn\\
H_{[321]}
\ee

$\bullet$ Level 7:
\be
H_7 = H_{[331]} = H_{[3211]} = H_{[31111]} = H_4 H_3\nn \\
H_{[61]} = H_{[43]} = H_{[2221]} = H_{[211111]} = H_{[21]} H_4\nn \\
H_{[52]} = H_{[22111]}^* \nn\\
H_{[511]} = H_{[421]} = H_{[322]} = H_{[1111111]} = H_4 H_{[111]} \nn \\
H_{[4111]}
\ee

$\bullet$ Level 8:
\be
H_8 = H_{[71]} = H_{[611]} = H_{[5111]} = H_{[44]} = H_{[431]}
= H_{[4211]} = H_{[41111]} = H_{[332]} = H_{[3221]} = \nn\\
 = H_{[311111]} = H_{[2222]} = H_{[2111111]} = H_{[11111111]} = H_4^2
\nn \\
H_{[62]} = H_{[53]} = H_{[22211]} = H_{[221111]} = H_4 H_{[22]} \nn \\
H_{[422]} = H_{[3311]}^* \nn \\
H_{[521]} = H_{[32111]}^*
\ee

\bigskip

\subsubsection{Towards colored $\hat{\cal A}$ polynomials
\label{refdif}}

Another way to represent these relations is to consider factorization of
the differences between HOMFLY in different representations,
for example
\be
H_{[41]} - H_{[21]}H_{[11]} \ \ \ \vdots \ \ \ \frac{\{q^3\}\{q^2\}}{\{q\}} &\Longleftarrow \ \
(\ref{Rpm})+(\ref{dir1r})+(\ref{fac21r6})
\ee
describes simultaneously several relations from above list.
%In the Appendix at the end of this paper
%we provide some more examples -- but they are rather material
%for further work on explicit universal formulas.
Reformulating (\ref{Rpm}) in such form can be the first step towards
derivation of ${\cal A}$-polynomial-like
equations \cite{Apols,IMMMfe,MMknots,Sulk,knotslast} for the colored knots.

In such differences there can be also factors containing $\{Aq^j\}$,
which imply additional relations at special values of $N$ -- they are also
of interest. We provide some examples  in the Appendix at the end of this paper.

\newpage

\section{
 Reduction of Kashaev polynomial to the special one }

Discussing knot polynomials at roots of unity
it is difficult to avoid looking at Kashaev polynomial \cite{Kap}
\be
 K_R^{\cal K}(A) = H_R^{\cal K} \left(q^2 = e^{2\pi i/|R|},A\right)
\ee
which is the value of colored HOMFLY at a primitive root of unity $q^{2|R|}=1$.
Doing so, we observe a remarkable fact:
for all single hook diagrams $R$ Kashaev polynomial
is easily expressed through the special polynomial:
\be
\boxed{
K_R^{\cal K}(A) = K^{\cal K}_{[1]}\left(A^{|R|}\right) =
H_{[1]}^{\cal K} \left(q^2=1,A^{|R|}\right)= \sigma_{[1]}^{\cal K}\left(A^{|R|}\right)
}
\ \ \ \ \ \ \forall  \  R=[r,1^k]
\label{Kashred}
\ee
It looks like an $A-q$ dual of the mysterious relation \cite{IMMMfe,Zhu} for
the Alexander polynomial $Al_R^{\cal K}(q)= H_R(A=1,q)$:
\be
Al_R^{\cal K}(q) = Al_{[1]}^{\cal K}\left(q^{|R|}\right)
\ \ \ \ \ \ \forall  \  R=[r,1^k]
\label{Alfact}
\ee
This means that Kashaev polynomial for single-hook diagrams -- and thus for all symmetric
representations, where it is mostly used -- is actually nothing more than the special one.
But, like Alexander, it becomes highly non-trivial whenever the number of hooks exceeds one.

%\section{Related properties of Kashaev polynomial}

\bigskip

{\footnotesize
For example, for torus knot ${\cal K} = [3,7]$
\be
K_{[1]}^{[3,7]} = \frac{5-16 A^2+12 A^4}{A^{16}} \nn \\ \nn \\
K_{[2]}^{[3,7]} = \frac{5-16 A^4+12 A^8}{A^{32}} = K_{[1]}^{[3,7]}(A^2) \nn \\
K_{[1,1]}^{[3,7]} = \frac{5-16 A^4+12 A^8}{A^{32}}  = K_{[1]}^{[3,7]}(A^2) \nn \\ \nn\\
K_{[3]}^{[3,7]} = \frac{5-16 A^6 +12 A^{12}}{A^{48}}   = K_{[1]}^{[3,7]}(A^3)\nn \\
K_{[2,1]}^{[3,7]} = \frac{5-16 A^6 +12 A^{12}}{A^{48}}  = K_{[1]}^{[3,7]}(A^3)\nn \\
K_{[1,1,1]}^{[3,7]} = \frac{5-16 A^6 +12 A^{12}}{A^{48}}  = K_{[1]}^{[3,7]}(A^3) \nn \\ \nn \\
K_{[4]}^{[3,7]} = \frac{5-16 A^8 +12 A^{16}}{A^{64}}  = K_{[1]}^{[3,7]}(A^4)\nn \\
K_{[3,1]}^{[3,7]} = \frac{5-16 A^8 +12 A^{16}}{A^{64}}  = K_{[1]}^{[3,7]}(A^4)\nn \\
K_{[2,2]}^{[3,7]} =
\frac{17+16A^2+20A^4+24A^6-40A^8-56A^{10}-68A^{12}-80A^{14}-24A^{16}}{A^{64}} \nn \\
K_{[2,1,1]}^{[3,7]} = \frac{5-16 A^8 +12 A^{16}}{A^{64}} = K_{[1]}^{[3,7]}(A^4) \nn \\
K_{[1,1,1,1]}^{[3,7]} = \frac{5-16 A^8 +12 A^{16}}{A^{64}} = K_{[1]}^{[3,7]}(A^4)\nn
\ee
}

\bigskip

For symmetric representations (\ref{Kashred}) can be reformulated as the property
of the differential expansion.
Eq.(\ref{diffe}) implies that Kashaev polynomial is a sum of just two terms:
\be
K_r^{\cal K}(A) = 1 +  \left.G_r^{\cal K}(A,q)
\cdot\{A/q\}\cdot \prod_{j=0}^{r-1} \{Aq^{r+j}\}\right|_{q=e^{i\pi/r}}
\ee
All other terms vanish, because $[r]=0$  at $q=e^{i\pi/r}$, what nullifies
binomial coefficients.

The product
\be
\left.\prod_{j=0}^{r-1} \{Aq^{r+j}\}\right|_{q=e^{i\pi/r}} = e^{- i\pi(r+1)/2} \{A^r\}
\ee
If ${\cal K}$ has {\it defect} zero \cite{KoModef}, i.e.
$G^{\cal K}_r\cdot \{A/q\} = F^{\cal K}_r\cdot \prod_{j=0}^{r-1}\{Aq^{j-1}\}$,
then we get the second product of the same kind and
\be
K_r^{\cal K}(A) = 1 + F_r\left(q=e^{i\pi/r},A\right) \cdot\{A^r\}^2 \ \ \ \ \ \ \ \ \
\text{provided}  \
\delta^{\cal K}=0
\ee
so that (\ref{Kashred}) means
\be
F_r\left(q=e^{i\pi/r},A\right)= F_1\left(q^2=1,A^r\right)
\ee
what is indeed true, for example, for $4_1$ with all $F_r=1$, for $3_1$ with
$F_r = (-)^r q^{- r(r-1)} A^{- 2r}$ and -- a little less trivially -- for other twist knots \cite{evo},
which all have defect zero.

For generic knot with arbitrary defect $\delta^{\cal K}$ eq.(\ref{Kashred}) implies that
\be
e^{- i \pi (r+1)/2 }\cdot G_r\left(q=e^{i\pi/r},A\right)\cdot\{Ae^{-i\pi/r}\}\ =\ G_1(q=1,A^r) \cdot \{A^r\}.
\label{GrKa}
\ee
Non-primitive roots of degree $2m$ are primitive of some lower degree,
thus the form of (\ref{Kashred}) and (\ref{GrKa}) is not universal for all the roots $q^{2m}=1$ --
in variance with (\ref{relsforG}).

Note in passing that since and $Al_{[1]}^{\cal K}(q=\pm 1) = 1$,
eq.(\ref{Alfact}) implies that
\be
\left. Al_{[r,1^k]}^{\cal K}(q) = 1 \ \right|_{q^{2(r+k)} =1}  \ \ \ \ \ \ \
\ee
i.e. Alexander polynomial is just trivial at the relevant root of unity --
and for primitive root this can be also considered as an implication of (\ref{Kashred})
at $A=1$.

\section{Conclusion}

In this paper we reported a number of interesting relations
for colored HOMFLY polynomial at the roots of unity,
which seem true for arbitrary knots.
They are obtained experimentally, by looking at explicit expressions,
implied by the recent \cite{gmmms,mmmrs} for a vast variety of knots.
Proofs are not yet available, and evidence can be not fully convincing,
especially for non-symmetric representations, where it comes from the
torus knots only.
Remarkable conjectures (\ref{Rpm}) and (\ref{Kashred}) cry for applying new
efforts to the study of colored polynomials -- what in the frame of \cite{mmmrs}
basically requires an effort in calculating Racah matrices.
A possible way to conceptual { proofs} can be within the Cherednik's DAHA
approach \cite{CheDAHA,Che,GoNe}, where something special also happens when $q$
is a root of unity \cite{Chepri}.
Even more distinguished are the roots of unity in the original method
of \cite{AgSha}.
As explained in sec.\ref{refdif} conjecture (\ref{Rpm}), if adequately reformulated,
is already sufficient to study the colored $\hat{\cal A}$-polynomials --
and this is another exciting direction, opened by our results.

\section*{Acknowledgements}

Our work is partly supported by grants NSh-1500.2014.2, RFBR 13-02-00478
and by the joint grants 15-52-50041-YaF, 14-01-92691-Ind-a, 15-51-52031-NSC-a.
Acknowledged is also support by  Brazilian Ministry of Science, Technology and Innovation
through the National Counsel of Scientific and Technological Development and
by Laboratories of Algebraic Geometry and Mathematical Physics, HSE.

\newpage

\section*{Appendix.
Possible germs of colored $\hat{\cal A}$-polynomials}
%\subsection*{
%Differences with factors $\{A/q^k\}$}

In this appendix we open a somewhat different line of factorization
properties, where what factors out from the differences between
colored polynomials are factors $\{A/q^k\}$ with some $k$.
When such factor appears, it means that the two HOMFLY polynomials
coincide when $A=q^{k}$.
Relations of this type can be equally important as those at roots of unity --
at least for the search of colored $\hat{\cal A}$-polynomials.

What we give is just a beginning of such list, and we do not formulate
any conjecture, comparable in generality to (\ref{Rpm}) --
this is left for the future.
\be
H^{\cal K}_{[21]}-H^{\cal K}_{[1]}   &\vdots &  \{A/q^2\}\{Aq^2\}\nn\\
H^{\cal K}_{[21]}-H^{\cal K}_{[3]}   &\vdots & \{A\}\{q^3\} \nn\\
H^{\cal K}_{[21]}-H^{\cal K}_{[111]}   &\vdots & \{A\}\{q^3\} \nn \\
\nn \\
H^{\cal K}_{[21]}-H^{\cal K}_{[32]}   &\vdots &   \{A/q^2\}\{Aq^4\} \nn\\
H^{\cal K}_{[21]}-H^{\cal K}_{[221]}   &\vdots &   \{Aq^2\}\{A/q^4\} \nn\\
H^{\cal K}_{[21]}-H^{\cal K}_{[321]}   &\vdots &    \{A/q^3\}\{A q^3\}
\ee

The next table is made closer in style to sec.\ref{Exa}:
instead of writing a factor $\{Aq^{-k}\}$ we list coincidences at $A=q^k$.

\bigskip

At $A=1$  relations are only between representations of the same size -- and
these were listed in sec.\ref{Exa}.

\bigskip

At $A=q$ all symmetric representations are trivial $H_{r}^{\cal K}(q=1,A)=1$
due to the factors $\{A/q\}$ in (\ref{diffe}),  additional relations
between our {\it reduced} (and thus non-vanishing) colored HOMFLY
are listed in the first column:

\bigskip

{\footnotesize
$$
\begin{array}{c|c|c|c|c|c}
A=q & A=q^2 & A=q^3 & A=q^4 & A=q^5 & \ \ \ldots \\
&&&&&\\
\hline
&&&&&\\
H^{\cal K}_{[2,2]} = H^{\cal K}_{[1,1,1]} &
H^{\cal K}_{[1]} = H^{\cal K}_{[2,1]} &
H^{\cal K}_{[1]} = H^{\cal K}_{[1,1]} &
H^{\cal K}_{[1]} = H^{\cal K}_{[1,1,1]}&
H^{\cal K}_{[1]} = H^{\cal K}_{[1,1,1,1]}
   \\ &&&&&  \\
H^{\cal K}_{[3,2]} = H^{\cal K}_{[2,1,1]}  &
H^{\cal K}_{[1]} = H^{\cal K}_{[3,2]} &
H^{\cal K}_{[1]} = H^{\cal K}_{[2,1,1]}&
H^{\cal K}_{[1]} = H^{\cal K}_{[2,1,1,1]}&
H^{\cal K}_{[1]} = H^{\cal K}_{[2,1,1,1,1]}
 \\ &&&&& \\
H^{\cal K}_{[3,3]} = H^{\cal K}_{[2,2,1]} &
H^{\cal K}_{[2]} = H^{\cal K}_{[3,1]} &
H^{\cal K}_{[1]} = H^{\cal K}_{[2,2,1]}&
H^{\cal K}_{[2]} = H^{\cal K}_{[2,2,2]}&
H^{\cal K}_{[1,1]} = H^{\cal K}_{[1,1,1]}
  \\ &&&&& \\
H^{\cal K}_{[3,3]} = H^{\cal K}_{[1,1,1,1]}   &
H^{\cal K}_{[2]} = H^{\cal K}_{[4,2]} &
H^{\cal K}_{[2]} = H^{\cal K}_{[2,2]}&
H^{\cal K}_{[2]} = H^{\cal K}_{[3,1,1,1]}&
H^{\cal K}_{[2,2]} = H^{\cal K}_{[2,2,2]}
\\  &&&&&\\
H^{\cal K}_{[4,2]} = H^{\cal K}_{[3,1,1]}      &
H^{\cal K}_{[3]} = H^{\cal K}_{[4,1]} &
H^{\cal K}_{[2]} = H^{\cal K}_{[3,1,1]}&
H^{\cal K}_{[1,1]} = H^{\cal K}_{[2,2,1,1]}&
H^{\cal K}_{[1,1,1]} = H^{\cal K}_{[1,1]}
      \\ &&&&& \\
H^{\cal K}_{[1,1,1]} = H^{\cal K}_{[2,2]}       &
H^{\cal K}_{[4]} = H^{\cal K}_{[5,1]} &
H^{\cal K}_{[3]} = H^{\cal K}_{[3,3]}&
H^{\cal K}_{[2,1]} = H^{\cal K}_{[2,2,1]}&
H^{\cal K}_{[2,1,1]} = H^{\cal K}_{[2,2,1,1]}
        \\ &&&&& \\
H^{\cal K}_{[2,1,1]} = H^{\cal K}_{[3,2]}        &
H^{\cal K}_{[1,1]} = H^{\cal K}_{[2,2]} &
H^{\cal K}_{[3]} = H^{\cal K}_{[4,1,1]}&
H^{\cal K}_{[1,1,1]} = H^{\cal K}_{[1]}&
H^{\cal K}_{[2,2,2]} = H^{\cal K}_{[2,2]}
          \\ &&&&& \\
H^{\cal K}_{[2,2,1]} = H^{\cal K}_{[3,3]}         &
H^{\cal K}_{[1,1]} = H^{\cal K}_{[3,3]} &
H^{\cal K}_{[1,1]} = H^{\cal K}_{[1]}&
H^{\cal K}_{[1,1,1]} = H^{\cal K}_{[2,1,1,1]}&
H^{\cal K}_{[1,1,1,1]} = H^{\cal K}_{[1]}
            \\ &&&&& \\
H^{\cal K}_{[2,2,1]} = H^{\cal K}_{[1,1,1,1]}      &
H^{\cal K}_{[2,1]} = H^{\cal K}_{[1]} &
H^{\cal K}_{[1,1]} = H^{\cal K}_{[2,1,1]}&
H^{\cal K}_{[2,2,1]} = H^{\cal K}_{[2,1]}&
H^{\cal K}_{[1,1,1,1]} = H^{\cal K}_{[2,1,1,1,1]}
              \\  &&&&&\\
H^{\cal K}_{[3,1,1]} = H^{\cal K}_{[4,2]}           &
H^{\cal K}_{[2,1]} = H^{\cal K}_{[3,2]} &
H^{\cal K}_{[1,1]} = H^{\cal K}_{[2,2,1]}&
H^{\cal K}_{[2,2,2]} = H^{\cal K}_{[2]}&
H^{\cal K}_{[2,2,1,1]} = H^{\cal K}_{[2,1,1]}
                \\ &&&&& \\
H^{\cal K}_{[3,2,1]} = H^{\cal K}_{[2,1,1,1]}  &
H^{\cal K}_{[2,2]} = H^{\cal K}_{[1,1]} &
H^{\cal K}_{[2,1]} = H^{\cal K}_{[3,2,1]}&
H^{\cal K}_{[2,2,2]} = H^{\cal K}_{[3,1,1,1]}&
H^{\cal K}_{[2,1,1,1,1]} = H^{\cal K}_{[1]}
                  \\ &&&&&\\
H^{\cal K}_{[1,1,1,1]} = H^{\cal K}_{[3,3]}     &
H^{\cal K}_{[2,2]} = H^{\cal K}_{[3,3]} &
H^{\cal K}_{[2,2]} = H^{\cal K}_{[2]}&
H^{\cal K}_{[2,1,1,1]} = H^{\cal K}_{[1]}&
H^{\cal K}_{[2,1,1,1,1]} = H^{\cal K}_{[1,1,1,1]}
                    \\ &&&&& \\
H^{\cal K}_{[1,1,1,1]} = H^{\cal K}_{[2,2,1]}    &
H^{\cal K}_{[3,1]} = H^{\cal K}_{[2]} &
H^{\cal K}_{[2,2]} = H^{\cal K}_{[3,1,1]}&
H^{\cal K}_{[2,1,1,1]} = H^{\cal K}_{[1,1,1]}&  \ldots
                      \\ &&&&& \\
H^{\cal K}_{[2,1,1,1]} = H^{\cal K}_{[3,2,1]}     &
H^{\cal K}_{[3,1]} = H^{\cal K}_{[4,2]} &
H^{\cal K}_{[3,1]} = H^{\cal K}_{[3,2]}&
H^{\cal K}_{[2,2,1,1]} = H^{\cal K}_{[1,1]}&
                        \\  &&&&& \\
H^{\cal K}_{[2,2,1,1]} = H^{\cal K}_{[1,1,1,1,1]}  &
H^{\cal K}_{[3,2]} = H^{\cal K}_{[1]} &
H^{\cal K}_{[3,2]} = H^{\cal K}_{[3,1]}&
H^{\cal K}_{[3,1,1,1]} = H^{\cal K}_{[2]}&
                          \\ &&&&& \\
H^{\cal K}_{[1,1,1,1,1]} = H^{\cal K}_{[2,2,1,1]}   &
H^{\cal K}_{[3,2]} = H^{\cal K}_{[2,1]} &
H^{\cal K}_{[3,3]} = H^{\cal K}_{[3]}&
H^{\cal K}_{[3,1,1,1]} = H^{\cal K}_{[2,2,2]}&
                            \\ &&&&& \\
\ldots & H^{\cal K}_{[3,3]} = H^{\cal K}_{[1,1]} &
H^{\cal K}_{[3,3]} = H^{\cal K}_{[4,1,1]}&\ldots &
\\ &&&&& \\
& H^{\cal K}_{[3,3]} = H^{\cal K}_{[2,2]} &
H^{\cal K}_{[1,1,1]} = H^{\cal K}_{[2,2,2]}&&
\\ &&&&&\\
& H^{\cal K}_{[4,1]} = H^{\cal K}_{[3]} &
H^{\cal K}_{[2,1,1]} = H^{\cal K}_{[1]}&&
\\ &&&&&\\
& H^{\cal K}_{[4,2]} = H^{\cal K}_{[2]} &
H^{\cal K}_{[2,1,1]} = H^{\cal K}_{[1,1]}&&
\\ &&&&&\\
&H^{\cal K}_{[4,2]} = H^{\cal K}_{[3,1]} &
H^{\cal K}_{[2,1,1]} = H^{\cal K}_{[2,2,1]}&&
\\ &&&&&\\
&H^{\cal K}_{[5,1]} = H^{\cal K}_{[4]} &
H^{\cal K}_{[2,2,1]} = H^{\cal K}_{[1]}&&
\\ &&&&&\\
&H^{\cal K}_{[2,2,2]} = H^{\cal K}_{[1,1,1,1]} &
H^{\cal K}_{[2,2,1]} = H^{\cal K}_{[1,1]}&&
\\ &&&&& \\
&H^{\cal K}_{[1,1,1,1]} = H^{\cal K}_{[2,2,2]} &
H^{\cal K}_{[2,2,1]} = H^{\cal K}_{[2,1,1]}&&
\\ &&&&& \\
      &\ldots&  H^{\cal K}_{[2,2,2]} = H^{\cal K}_{[1,1,1]}&&
\\ &&&&& \\
&&H^{\cal K}_{[3,1,1]} = H^{\cal K}_{[2]}&&
 \\ &&&&& \\
&&H^{\cal K}_{[3,1,1]} = H^{\cal K}_{[2,2]}&&
 \\ &&&&& \\
&&H^{\cal K}_{[3,2,1]} = H^{\cal K}_{[2,1]}&&
  \\ &&&&& \\
&&H^{\cal K}_{[4,1,1]} = H^{\cal K}_{[3]}&&
 \\ &&&&& \\
&&H^{\cal K}_{[4,1,1]} = H^{\cal K}_{[3,3]}&&
\\ &&&&& \\
       &&\ldots&&&
\end{array}
$$
}

\end{document}